\documentclass{article}

\usepackage{arxiv}

\usepackage[utf8]{inputenc} 
\usepackage[T1]{fontenc}   
\usepackage{hyperref}    
\usepackage{url}           
\usepackage{booktabs}      
\usepackage{amsfonts}       
\usepackage{nicefrac}    
\usepackage{microtype}      
\usepackage{lipsum}
\usepackage{changepage}
\usepackage{graphicx}
\usepackage{amsmath}
\usepackage{amsfonts}
\usepackage{amssymb}
\usepackage{makeidx}
\usepackage{algorithm,algorithmic}
\usepackage[switch]{lineno}
\usepackage{xcolor}
\usepackage{bm}
\usepackage[justification=centering]{caption}
\usepackage{amsfonts}
\usepackage[bbgreekl]{mathbbol}
\DeclareSymbolFontAlphabet{\mathbbm}{bbold}
\DeclareSymbolFontAlphabet{\mathbb}{AMSb}%
\usepackage{amsthm}

\usepackage[title]{appendix}%

\title{Predictive Controlled Music}

\author{
  Midhun T. Augustine \\
  University of Alberta, Canada \\
  \texttt{midhun@ualberta.ca} \\
 \vspace{.01cm}\\
    25 - 12 - 2025\\ \vspace{.01cm}}

\begin{document}
\maketitle

\begin{abstract}
This paper presents a new approach to algorithmic composition, called predictive controlled music (PCM), which combines model predictive control (MPC) with music generation. PCM uses dynamic models to predict and optimize the music generation process, where musical notes are computed in a manner similar to an MPC problem by optimizing a performance measure. A feedforward neural network–based assessment function is used to evaluate the generated musical score, which serves as the objective function of the PCM optimization problem. Furthermore, a recurrent neural network model is employed to capture the relationships among the variables in the musical notes, and this model is then used to define the constraints in the PCM. Similar to MPC, the proposed PCM computes musical notes in a receding-horizon manner, leading to feedback controlled  prediction. Numerical examples are presented to illustrate the PCM generation method.
\end{abstract}


\section{Introduction}
\par Electronic music generation involves the composition and synthesis of music through the use of electronic instruments and software \cite{Loy2006,Hiller1959}. A  promising area within this field is algorithmic composition, also referred to as automated music generation \cite{Nierhaus2009}. Algorithmic composition relies on mathematical models to predict or generate musical pieces in an automated manner. Early approaches in this domain primarily employed predictive models such as autoregressive models, Markov chains, and Monte Carlo methods \cite{Loy2006,Nierhaus2009}. More recently, with the rapid development of data science, AI-based modeling and prediction have emerged as a prominent and increasingly active area of research \cite{McDonald2015,Russell2021}.
Deep learning (DL), a subfield of AI, focuses specifically on constructing and training deep neural networks (NNs) \cite{Russell2021,Goodfellow2016}.  DL is capable of modeling static relationships through feedforward NN \cite{McCulloch1943,Cybenko1989} and autoencoder \cite{Kramer1991} architectures, and dynamic relationships through recurrent NN (RNN) \cite{Hopfield1984,Narendra1990}, transformers \cite{Vaswani2017}, and related sequence models. Consequently, DL offers a generalized framework capable of representing complex patterns and relationships. One rapidly expanding application area of DL is content generation, which includes text, images, audio, and video \cite{Goodfellow2016}. Within this domain, music generation using DL techniques has been studied extensively \cite{Briot2020}.
\par Initial research on NN–based music generation primarily employed feedforward NN architectures \cite{Hild1991,Bresin1998}. Subsequently, RNN–based models were introduced, as they effectively capture temporal relationships in music and are therefore more suitable for automated generation  \cite{Lewandowski2012,Oore2018,Jaques2016}. A RNN-based probabilistic model is introduced in \cite{Lewandowski2012} for learning and predicting temporal relationships in high-dimensional sequences, with a particular focus on polyphonic music. The performance-RNN model \cite{Oore2018} is designed to learn and generate expressive musical performances. 
In \cite{Jaques2016}, the authors present a sequence tutor designed to enhance the quality and structure of generated sequences; this approach employs an RNN trained via reinforcement learning to improve performance, and its effectiveness for music generation is demonstrated through numerical experiments. The application of unsupervised feature-learning methods to the modeling of expressive dynamics is examined in \cite{Grachten2014,Engel2017}. Specifically, \cite{Grachten2014} employs learned features to predict note intensities, while \cite{Engel2017} proposes a wavenet autoencoder capable of interpolating timbre to generate new types of realistic sounds. In \cite{Lattner2018}, a convolutional NN–based generative model is proposed to impose higher-level structural and tonal properties on polyphonic music.  In \cite{Eck2002,Sturm2016}, long short-term memory (LSTM)-based dynamic models are employed to extract temporal structures and music transcription models that are useful for specific contexts of music composition. Furthermore, \cite{Malik2017} introduces an LSTM  model capable of learning to perform musical scores. 
Although several approaches for DL based music generation have been proposed in the past, the area of controlled music generation is less explored. Most of the existing approaches focus on predictive models for generation \cite{Briot2020}.

\par   The current work focuses on controlled music generation, which combines concepts from control theory into algorithmic composition. 
Amongst the modern control approaches, the model predictive control (MPC) is a popular approach that computes the control sequence by optimizing a performance measure computed using model-based prediction \cite{Borrelli2017}. Even though MPC is applied in a wide range of areas such as industrial processes, automated vehicles, finance, building automation, and biological systems, among others \cite{Mayne2014}, the application in music generation is not explored yet, to the best of the author's knowledge. 
This motivates the proposed predictive controlled music (PCM) approach, where the MPC principles are used for automated score generation. An An RNN-based model is used in PCM to capture the relationships among the variables that describe the properties of the musical piece. Further, an assessment function based on an NN-based rating function and aesthetic quality measure is used for evaluating the score generated. The proposed PCM generates the optimal score sequence by maximizing the assessment function, subject to constraints defined using the RNN model. This distinguishes it from the existing AI-based approaches, where the control aspect is not incorporated, and generation is through prediction only.

\par The rest of the paper is organized as follows. Section 2 briefly discusses the preliminary concepts from music generation, MPC, and RNN. Section 3 presents the PCM approach, which uses an RNN model-based predictive control scheme for music generation.   Section 4 illustrates the numerical implementation of the proposed PCM approach. 
Finally, section 5 contains the conclusions and future directions of the paper.

\section{Preliminaries}
\subsection{Music generation}
\label{secmusicgen}
\begin{figure}[H]
 		\begin{center}
 		\includegraphics [scale=0.18] {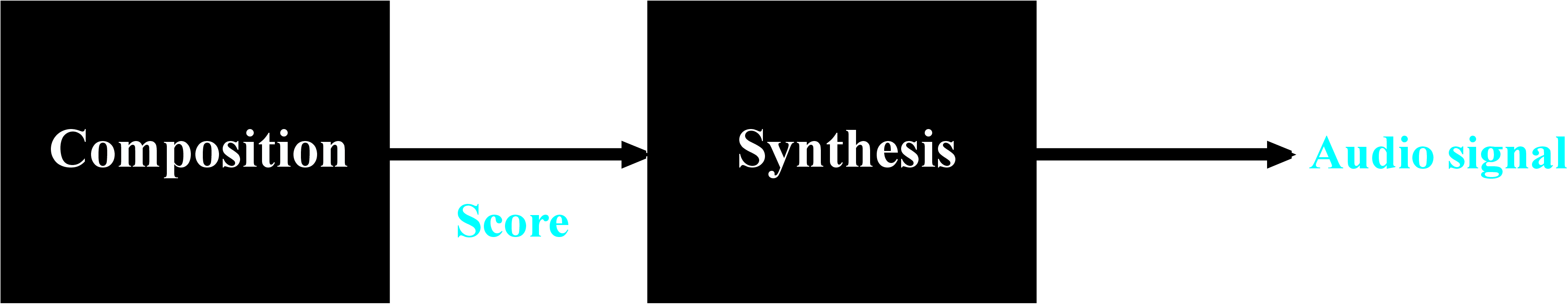}
 		\caption{{Music generation: simplified block diagram.}}
   \label{figmusicgen}
 	\end{center}
 \end{figure}

In this paper, we focus on electronic music generation, which is associated with the composition of a score and the synthesis of an audio signal based on the score.  This overall process is illustrated in Fig. \ref{figmusicgen}.  Key terminology relevant to music generation is summarized in Table~\ref{table:1}, primarily based on the definitions provided in \cite{Loy2006}. 
A fundamental property of a musical signal is frequency, which characterizes the behavior of the sound wave.
In music, frequency is commonly represented in terms of pitch, defined as the logarithmic ratio of a given frequency to a reference frequency:
\begin{equation}
    \label{eqpitch}
    p=c \hspace{0.2cm} log_{2}(\frac{f}{f_{0}})
\end{equation}
where $f_{0}$ is the reference frequency and $c$ is the scaling constant; $c=12$ gives the pitch in semitones, whereas $c=1$ yields pitch in octaves.
More recently, frequency is also represented using the Musical Instrument Digital Interface (MIDI) number, which is related to pitch by:
\begin{equation}
m = m_{0} + p
\end{equation}
where $m_{0}$ is the MIDI number corresponding to the reference frequency $f_{0}$. The formula for obtaining frequency in Hz from pitch and MIDI number can be obtained by inverting Eq. (\ref{eqpitch}) as \cite{Puckette2007}:
\begin{equation}
    \label{eqfreq}
    f=f_{0}2^{\frac{p}{c}}=f_{0}\times 2^{\frac{m-m_{0}}{c}}
\end{equation}
\begin{table}[h!]
\caption{Terminologies in music generation.}
\centering
\begin{tabular}{| c  c  p{12cm} |} 
 \hline
 Term & Notation & Definition \\ [0.5ex]
 \hline\hline

 Pitch & $p$ & 
 Pitch is a fundamental property of a sound wave that is determined by its frequency $f$.  
 Pitch refers specifically to perceived sound within the human hearing range (20 Hz to 20,000 Hz) and is normally represented in a logarithmic scale, whereas frequency is a general physical property that can also describe vibrations outside this range.
 \\[1.5ex]
Velocity & $v$ & Loudness of sound is quantified as velocity, which typically indicates how fast a key is pressed in a piano or how hard a note is played. Generally, velocity is multiplied by the synthesized sound to scale its amplitude.
 \\[1.5ex]
 Tempo & $q$ & Tempo (measured in beats per minute (BPM)) refers to the overall speed at which a musical piece is performed. Tempo defines the pace of a song; for example, in a remix, the tempo is often increased to create a faster, more energetic feel. 
  \\[1.5ex]
Onset & $s$ & 
 Onset denotes the starting time of the sound.
\\[1.5ex]
 Duration & $d$ & 
 Duration refers to the length of time over which a sound is perceived.
 \\[1.5ex]
Note & $\textbf{z}$ & Note is characterized by pitch, velocity,  onset and duration: $\textbf{z}_{k}=\left[\begin{matrix} p_{k} & v_{k} & s_{k} & d_{k}\end{matrix}\right]$.
 \\[1.5ex]
 Score & $\textbf{Z}$ & Score is an ordered sequence of notes, i.e., $\textbf{Z}=\left[\begin{matrix} \textbf{z}_{1} & \textbf{z}_{2} & \dots & \textbf{z}_{\text{N}_{s}}\end{matrix}\right]$
  \\[1.5ex]
  Interval & $\varDelta p$ & Interval is defined as the difference in pitch between two notes: $\varDelta p_{k}=p_{k}-p_{k-1}$. Commonly measured in semitones or octaves.
 \\[1.5ex]
 \hline
\end{tabular}
\label{table:1}
\end{table}
In standard MIDI representation, the reference frequency is chosen as $f_{0}=440$ Hz, the corresponding MIDI number reference is $m_{0}=69$.
 For example, substituting $c=12$ and $f=f_{0}=440$ ($m_{0}=69$) in Eq. (\ref{eqpitch}) gives the pitch as zero ($m=69$) and $f=2f_{0}=880$ Hz corresponds to a pitch of 12 semitones ($m=69+12=81$). Another important property of sound is timbre, which refers to the spectral characteristics that allow us to distinguish between different instruments or voices, even when they produce the same pitch. Timbre can be characterized using spectral analysis techniques, such as the Fourier transform. In general, pitch, tempo, timbre, onset, and duration describe the temporal and spectral features of a sound wave, while velocity reflects the perceived intensity of the sound. The variation of velocity or loudness over time is referred to as dynamics in music terminology.  A melody is the sequential organization of notes over time, whereas harmony refers to the simultaneous combination of two or more notes. Composition refers to the creation of a musical score, where a sequence of musical notes containing elements such as pitch, velocity, onset, and duration is defined.  This paper mainly focuses on the composition stage, where the primary contribution is generating a score or sequence of note vectors:
\begin{equation}
    \label{eqscore}\textbf{Z}=\left[\begin{matrix} \textbf{z}_{1} & \textbf{z}_{2} & \dots & \textbf{z}_{\text{N}_{s}}\end{matrix}\right]
\end{equation}
using PCM, where $\textbf{z}_{k}=\left[\begin{matrix} p_{k} & v_{k} & s_{k} & d_{k}\end{matrix}\right]^{\top}, k=1,\dots,\text{N}_{s}$ and $\text{N}_{s}$ is the number of note vectors generated or the simulation time for PCM.
\par The synthesis stage involves generating a sound waveform from a musical score as in Eq. (\ref{eqscore}). This can be accomplished by performing the score using various instruments and recording the resulting audio. Alternatively, electronic synthesis can be used, in which the sound waveform corresponding to the specified pitch and velocity is generated algorithmically. 
 In this paper, a very basic electronic synthesis method is employed, in which a sine wave is generated at the frequency given in Eq. (\ref{eqfreq}), corresponding to the pitch or MIDI note number. A piano sound wave based on the note frequency $f$ is then generated using the following equation, which models the piano spectrum:
\begin{equation}
x(t) = \sum_{k=1}^{N_{h}} a_k \sin(2\pi k f t)
\end{equation}
where $t \in \mathbb{R}^{+}$ denotes continuous time and $N_{h}$ is the number of harmonic terms included in the signal, which decides its timbre.  The resulting sound signal is sampled at a high frequency, $f_{s} = 44{,}100$ Hz (with sampling period $T_{s} = \frac{1}{44{,}100} \approx 22.7~\mu\text{s}$), to produce a digital audio signal that can be stored as a vector.  The resulting waveform is further processed using various filters and wave-shaping algorithms to improve the perceptual quality of the synthesized sound. A low-pass filter with a cutoff frequency of 3 kHz is applied to produce a piano-like tone. In addition, a short exponentially decaying noise component is added to the generated signal to simulate the hammer impact. Furthermore, an envelope function such as the attack, decay, sustain, and release (ADSR) envelope is used to shape the amplitude of the signal over each note’s duration by multiplying the waveform with the ADSR output. The ADSR function can be defined using either linear or exponential segments. The linear ADSR function is defined as \cite{Puckette2007}:
\begin{equation}
    \label{eqadhr}
    f_{ADSR}(t)=\begin{cases}
\frac{t}{t_{A}} \hspace{0.3cm} & \text{for} \hspace{0.15cm} 0\leq t\leq t_{A}\\
1-(1-S)\frac{t-t_{A}}{t_{D}} \hspace{0.3cm} &\text{for} \hspace{0.15cm} t_{A}<t \leq t_{A}+t_{D}\\
S \hspace{0.3cm} & \text{for} \hspace{0.15cm} t_{A}+t_{D} < t\leq t_{A}+t_{D}+t_{S}\\
S-S\frac{t-(t_{A}+t_{D}+t_{S})}{t_{R}} \hspace{0.3cm} & \text{for} \hspace{0.15cm} t_{A}+t_{D}+t_{S} < t\leq t_{A}+t_{D}+t_{S}+t_{R}
\end{cases}
\end{equation}
where $0<S<1$ is the sustain value,  $t_{A},t_{D},t_{S},t_{R}$ are the attack, decay, sustain, and release durations for which the sum equals the note duration: $d=t_{A}+t_{D}+t_{S}+t_{R}$.

\subsection{Model Predictive Control}
\label{secmpc}
Model predictive control (MPC) is a modern control strategy that employs model-based optimization \cite{Borrelli2017}. In MPC, a model is used to predict the system’s output over a predefined horizon. The control input sequence over a finite horizon is then determined by optimizing the predicted behavior, which is typically defined by a cost function. In this paper, the horizon for prediction and control is chosen to be the same and denoted by $\text{N}$. The model can be represented in various forms, such as linear or nonlinear, input-output or state-space representations. The corresponding MPC formulations can be found in \cite{Augustine2025}. This paper focuses on the use of a nonlinear input-output model, as described below:
\begin{equation}
    \label{eqnlfy} 
    \mathbf{y}_{k+1}=\mathbf{f}(\mathbf{y}_{k},\mathbf{u}_{k})
\end{equation}
where $k\in \mathbb{Z}^{+}$, $\mathbf{y}_{k}\in \mathbb{Y} \subseteq \mathbb{R}^{p}$ ,  $\mathbf{u}_{k}\in \mathbb{U} \subseteq \mathbb{R}^{m}$, $\mathbf{f}:\mathbb{Y} \times \mathbb{U} \rightarrow \mathbb{Y}$ is a nonlinear mapping. The predicted input and output sequence for MPC  can be defined as:
\begin{equation}
\label{eqyprednl}
\mathbf{U}_{k}=\left[\begin{matrix} \mathbf{u}_{k|k} \\  \vdots \\ \mathbf{u}_{k+\mathrm{N}-1|k}\end{matrix}\right], \hspace{0.5cm}\mathbf{Y}_{k+1}=\left[\begin{matrix} \mathbf{y}_{k+1|k} \\  \vdots \\ \mathbf{y}_{k+\mathrm{N}|k}\end{matrix}\right]=\left[\begin{matrix} \mathbf{f}(\mathbf{y}_{k|k},\mathbf{u}_{k|k})  \\  
 \vdots \\  
\mathbf{f}(\dots\mathbf{f}(\mathbf{y}_{k|k},\mathbf{u}_{k|k}), \mathbf{u}_{k+\mathrm{N}-1|k})
\end{matrix}\right]= \mathbf{f}_{\mathrm{yp}}(\mathbf{y}_{k|k},\mathbf{U}_{k})
\end{equation}
where $\mathbf{u}_{i|k}$ and $\mathbf{y}_{i|k}$ denotes the input and output at time instant $i$ computed at time instant $k$. 
Using this, the cost function for output-based MPC is defined as:
\begin{equation}
\label{eqmpcjk}
\begin{aligned}
    J&=\sum_{i=k}^{k+\mathrm{N}-1}[\mathbf{y}_{\mathrm{r}}-\mathbf{y}_{i+1|k}]^{\top}\mathbf{Q}[\mathbf{y}_{\mathrm{r}}-\mathbf{y}_{i+1|k}]+[\mathbf{u}_{\mathrm{r}}-\mathbf{u}_{i|k}]^{\top}\mathbf{R}[\mathbf{u}_{\mathrm{r}}-\mathbf{u}_{i|k}] \\
    &=[\mathbf{Y}_{\text{r}}-\mathbf{Y}_{k+1}]^{\top}\mathbf{Q}_{\mathbf{Y}} [\mathbf{Y}_{\text{r}}-\mathbf{Y}_{k+1}] + [\mathbf{U}_{\text{r}}-\mathbf{U}_{k}]^{\top} \mathbf{R}_{\mathbf{U}}[\mathbf{U}_{\text{r}}-\mathbf{U}_{k}]
    \end{aligned}
\end{equation}
where  $\mathbf{Q}\in \mathbf{R}^{p\times p},$ $\mathbf{R}\in \mathbf{R}^{m\times m},$ are the output and input weighting matrices, $\mathbf{Q}_{\textbf{Y}}=diag(\textbf{Q},\textbf{Q},\dots,\textbf{Q})\in \mathbf{R}^{p\text{N}\times p\text{N}},$ $\mathbf{R}_{\textbf{U}}=diag(\textbf{R},\textbf{R},\dots,\textbf{R})\in \mathbf{R}^{m\text{N}\times m\text{N}}$ are the block diagonal weighting matrices for the predicted output and input sequences, $\textbf{y}_{\text{r}}\in \mathbb{R}^{p}$, $\textbf{u}_{\text{r}}\in \mathbb{R}^{m}$ are the output and input references, and $\textbf{Y}_{\text{r}}=\in\mathbf{R}^{p\text{N}}, \textbf{U}_{\text{r}}\in\mathbf{R}^{m\text{N}}$ are the reference sequence contains output and input references over the prediction horizon. In this paper, fixed weighting matrices are employed; however, the approach can be extended to incorporate time-varying weighting matrices that assign greater weight to the terminal outputs (predicted outputs towards the end of the horizon).
Then, an output-based nonlinear MPC problem 
for the nonlinear system (\ref{eqnlfy}) with the current output $\mathbf{y}_{k|k}$, can be formulated as the following nonlinear programming problem (NLP):
\begin{equation}
\label{eqnmpc}
    \begin{aligned}    \underset{\mathbf{U}_{k},\mathbf{Y}_{k+1}}{\inf} \hspace{.2cm} &J\\ \mathrm{subject \hspace{0.1cm}to}
    ~&~  \mathbf{Y}_{k+1}\in \mathbb{Y}^{\mathrm{N}},\hspace{0.2cm} \mathbf{U}_{k} \in \mathbb{U}^{\mathrm{N}}\\
    ~&~ \mathbf{y}_{i+1|k}=\mathbf{f}(\mathbf{y}_{i|k},\mathbf{u}_{i|k}) \hspace{0.7cm}k\in \mathbb{Z}^{+},i=k,...,k+\mathrm{N}-1.
    \end{aligned}
\end{equation}

The predicted output sequence $\mathbf{Y}_{k+1}$ and control sequence $\mathbf{U}_{k}$ in Eq. (\ref{eqyprednl}) can be used to 
rewrite the optimization problem for output-based nonlinear MPC as:
\begin{equation}
\label{eqnmpc2y}
\begin{aligned}
 \underset{\mathbf{U}_{k}}{\inf}  ~&~  [\mathbf{Y}_{\text{r}}-\mathbf{f}_{\mathrm{yp}}(\mathbf{y}_{k|k},\mathbf{U}_{k})]^{\top}\mathbf{Q}_{\mathbf{Y}} [\mathbf{Y}_{\text{r}}-\mathbf{f}_{\mathrm{yp}}(\mathbf{y}_{k|k},\mathbf{U}_{k})] + [\mathbf{U}_{\text{r}}-\mathbf{U}_{k}]^{\top} \mathbf{R}_{\mathbf{U}}[\mathbf{U}_{\text{r}}-\mathbf{U}_{k}] \\
\mathrm{subject \hspace{0.1cm}to} ~&~ \mathbf{f}_{\mathrm{yp}}(\mathbf{y}_{k|k},\mathbf{U}_{k}) \in \mathbb{Y}^{\mathrm{N}}, \hspace{0.2cm}
\mathbf{U}_{k} \in \mathbb{U}^{\mathrm{N}}.
     \end{aligned}
\end{equation}
The cost function and constraint involve the nonlinear mapping $\mathbf{f}_{\mathrm{yp}}$ which makes Eq. (\ref{eqnmpc2y}) an NLP.
The MPC strategy is implemented by solving this NLP at each time instant and applying the first component of $\mathbf{U}_{k}^{*}$ as the control input.

\subsection{Recurrent Neural Network}
\begin{figure}[H]
 		\begin{center}
 		\includegraphics [scale=0.7] {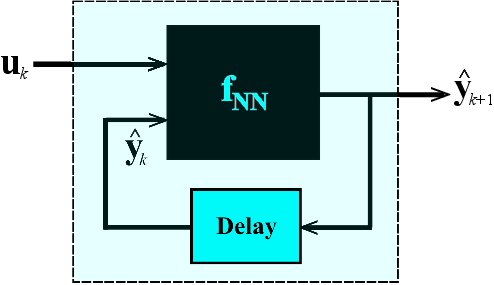}
 		\caption{{RNN Block diagram.}}
   \label{figrnn}
 	\end{center}
 \end{figure}

Recurrent Neural Networks (RNNs) are dynamic models used to represent nonlinear temporal relationships within data \cite{Hopfield1984,Narendra1990}. 
 Unlike feedforward NNs, RNNs are bi-directional NNs where the data can flow in the backward direction as well through recurrent (feedback) connections, as shown in Fig. \ref{figrnn}.
RNNs can be used to model systems described by the equation in Eq. (\ref{eqnlfy}), where the structure and parameters of the function $\textbf{f}$ are unknown. The objective of RNN is to approximate the $\textbf{f}$ in Eq. (\ref{eqnlfy}).  For that, the available information is used, which is the  training data that consists of input and output vectors: 
    \begin{equation}
      \label{equy}  
     \textbf{U}_{\textbf{d}}=\left[\begin{matrix} \textbf{u}_{1} & \textbf{u}_{2} & \dots & \textbf{u}_{\text{D}}\end{matrix}\right], \hspace{0.5cm}
     \textbf{Y}_{\textbf{d}}=\left[\begin{matrix} \textbf{y}_{1} & \textbf{y}_{2} & \dots & \textbf{y}_{\text{D}}\end{matrix}\right]
    \end{equation}
    where $\textbf{u}_{k}$ and $\textbf{y}_{k},$ $k=1,\dots, \text{D}$ are the input and output samples at $\text{D}$ time instants. The input and output vectors are considered to be generated by the actual system, which can be represented as in Eq. (\ref{eqnlfy}). Then RNNs approximate the system in Eq. (\ref{eqnlfy}) as:
\begin{equation} 
\label{eqyhatrnn}
\hat{\textbf{y}}_{k+1}=\textbf{f}_{\text{NN}}(\hat{\textbf{y}}_{k},\textbf{u}_{k};\bm{\theta}_{\text{f}})=\textbf{f}_{\text{L}}(\textbf{f}_{\text{L-1}}(\dots(\textbf{f}_{1}(\hat{\textbf{y}}_{k},\textbf{u}_{k})))).
\end{equation}
where $\bm{\theta}_{\text{f}}\in \mathbb{R}^{n_f}$ contains parameters of RNN, which are the weights and biases of all layers.
 
During training, the objective is to find the optimal parameters that best fit the given data in Eq. (\ref{equy}). For a given training data as in Eq. (\ref{equy}), both the actual and predicted output sequences can be constructed as follows:
\begin{equation}
 \label{eqyrnn}
\textbf{Y}=\left[\begin{matrix} \textbf{y}_{1}\\ \textbf{y}_{2} \\ \vdots \\ \textbf{y}_{\text{D}}\end{matrix}\right],
\hspace{0.5cm}
\hat{\textbf{Y}}=\left[\begin{matrix} \textbf{y}_{1} \\  
\textbf{f}_{\text{NN}}(\textbf{y}_{1},\textbf{u}_{1})  \\ \vdots \\  
\textbf{f}_{\text{NN}}(\dots\textbf{f}_{\text{NN}}(\textbf{y}_{1},\textbf{u}_{1}), \textbf{u}_{\text{D}-1})
\end{matrix}\right]=\mathbf{f}_{\mathrm{yp}}(\bm{\theta}_{\text{f}}).
\end{equation}
The loss function for the RNN training is chosen as the output prediction error:
\begin{equation}
\label{eqjyrnn}
L=\sum_{k=1}^{\text{D}}  {\parallel \mathbf{y}_{k} - \hat{\mathbf{y}}_{k} \parallel }^{2}=[\textbf{Y}-\mathbf{f}_{\mathrm{yp}}(\bm{\theta}_{\text{f}})]^{\top}[\textbf{Y}-\mathbf{f}_{\mathrm{yp}}(\bm{\theta}_{\text{f}})]
\end{equation}
 This gives the optimization problem for RNN:
\begin{equation}
\label{eqrnnloss}
\underset{\bm{\theta}_{\text{f}}}{\min} \hspace{.2cm}  L.
\end{equation} 
Major variants of RNNs proposed in the literature include long short-term memory (LSTM) network \cite{Hochreiter1997} and gated recurrent unit (GRU) \cite{Cho2014}.
\section{Predictive Controlled Music (PCM)}
\label{secpcm}

This section presents the proposed predictive controlled music (PCM) for algorithmic composition. The main objective of the PCM
is to optimize the music generation to create realistic musical pieces. To achieve this, the PCM uses  MPC where the musical notes are generated by solving an MPC problem similar to Eq. (\ref{eqnmpc}). A number of variables, such as pitch, velocity, tempo, etc., can be selected as input and output variables for the PCM. Further, the PCM uses an RNN model, which captures the relationships between input and output variables. For the PCM implementation considered in this paper, pitch is used as the control input and velocity is used as the output. Using the training data consists of samples of pitch and velocity, an RNN model, as in Eq. (\ref{eqyhatrnn}), is trained, which is used as the model in PCM.   The predicted output sequence $\textbf{Y}_{k+1}$ and the input sequence $\textbf{U}_{k}$ for PCM are defined as in Eq. (\ref{eqyprednl}), with the RNN function $\textbf{f}_{\text{NN}}$ in Eq. (\ref{eqyhatrnn}) is used in place of the nonlinear mapping $\textbf{f}$ in Eq. (\ref{eqnlfy}). In the proposed PCM, an assessment function is used instead of the loss function $J$ in Eq. (\ref{eqmpcjk}). The assessment function $A(\textbf{U}_{k},\textbf{Y}_{k+1})$ characterizes the quantitative and qualitative aspects of the score generated. Then, in PCM, the objective is to maximize the assessment value of the score generated, i.e., maximize the score quality.
The resultant optimization problem for PCM becomes:
\begin{equation}
\label{eqnpcm}
    \begin{aligned}    \underset{\mathbf{U}_{k},\mathbf{Y}_{k+1}}{\inf} \hspace{.2cm} &-A\\ \mathrm{subject \hspace{0.1cm}to}
    ~&~  \mathbf{Y}_{k+1}\in \mathbb{Y}^{\mathrm{N}},\hspace{0.2cm} \mathbf{U}_{k} \in \mathbb{U}^{\mathrm{N}}\\
    ~&~ \mathbf{y}_{i+1|k}=\mathbf{f}_{\text{NN}}(\mathbf{y}_{i|k},\mathbf{u}_{i|k}) \hspace{0.7cm}k\in \mathbb{Z}^{+},i=k,...,k+\mathrm{N}-1.
    \end{aligned}    
\end{equation}
In this paper, a simple choice for $A$ is considered where $A$ is defined as the sum of an aesthetic quality function $q$ and a score rating function $s$ as:
\begin{equation}
    \label{eqq}
    A=s(\textbf{U}_{k})+ \alpha \hspace{0.1cm} q(\textbf{U}_{k},\textbf{Y}_{k+1})
\end{equation}
where $q:\mathbb{U}^{\text{N}}\times \mathbb{Y}^{\text{N}}\rightarrow \mathbb{R}$ is the function representing the aesthetic quality of the score and $s:\mathbb{U}^{\text{N}}\rightarrow \mathbb{R}$ is the score rating function which is approximated using a feedforward NN:
\begin{equation}
\label{eqqnn}
   s(\mathbf{U}_{k})= s_{\text{NN}}(\textbf{U}_{k};\bm{\theta}_{\text{s}})=\textbf{f}_{\text{L}}(\textbf{f}_{\text{L-1}}(\dots(\textbf{f}_{1}(\textbf{U}_{k}))))
\end{equation}
The parameters of the rating function are obtained by training an NN defined as in Eq. (\ref{eqqnn}). The training dataset is constructed by manually rating 100 different musical scores, with each score assigned a rating between 0 and 1. Around $15\%$ of the scores is selected such that the same note is repeatedly played; these scores are assigned a rating of zero. Among the 100 scores, fifty correspond to well-established songs such as Twinkle Twinkle Little Star, Ode to Joy, each of which is assigned a rating of 1. The remaining scores consist of slightly modified versions of these songs or randomly generated note sequences, and are assigned ratings between 0 and 1 based on subjective evaluation after listening to the corresponding musical pieces. It is important to note that the labeling used for the rating function is subjective and may vary across individuals. The realism of the rating function can be improved by increasing the size of the training dataset and incorporating feedback from a large number of users, with the mean rating used as the final label. In general, to the best of the authors’ knowledge, no standardized rating system exists for music analogous to systems such as IMDb for movies. The training data used for the rating NN can be represented as:
  \begin{equation}
      \label{eqratinguy}  
\textbf{U}_{\textbf{s}}=\left[\begin{matrix} \textbf{U}_{s_1} & \textbf{U}_{s_2} & \dots & \textbf{U}_{s_\text{H}}\end{matrix}\right], \hspace{0.5cm}
     \textbf{S}=\left[\begin{matrix} s_1 & s_2 & \dots & s_\text{H}\end{matrix}\right]
    \end{equation}
    where $\textbf{U}_{s_i}\in \mathbb{U}^{\text{N}},$ $i=1,\dots,\text{H}$ are the training input samples where each sample represents a distinct score of length $\text{N}$, and $\text{H}$ is the number of samples.
The loss function for the rating NN is chosen as:
\begin{equation}
\label{eqlossrating}
L_{s}=\sum_{i=1}^{\text{H}}(s_i-s_{\text{NN}}(\textbf{U}_{s_i};\bm{\theta}_{\text{s}}))^{2}.
\end{equation}
This gives the optimization problem for rating NN:
\begin{equation}
\label{eqratingopt}
\underset{\bm{\theta}_{\text{s}}}{\min} \hspace{.2cm}  L_{s}.
\end{equation} 
The aesthetic quality of the score can be evaluated in different ways \cite{Birkhoff1933,Jin2023}. One popular measure is  Birkhoff’s aesthetic measure, which is used in the proposed PCM and is defined as \cite{Birkhoff1933}:
\begin{equation}
\label{eq_birkhoff}
        q = \frac{O}{C}.
\end{equation}
where $O$ and $C$ represent the order and complexity of the song.  Higher values of $q$ correspond to melodies with more ordered structure, while lower values indicate more irregular or highly complex 
structures. Let $\Delta_{u} = \{\varDelta p_{1}, \varDelta p_{2}, \ldots, \varDelta p_{\text{N}}\}$ be the sequence of melodic intervals 
between successive pitches in $\textbf{U}_{k}$ and $\Delta_{y} = \{\varDelta v_{1}, \varDelta v_{2}, \ldots, \varDelta v_{\text{N}}\}$ be the difference successive velocities in $\textbf{Y}_{k+1}$. Define $h_{u_i}$ as the histogram count of pitch interval 
type $i$ and $h_{y_i}$ as the histogram count for the velocity interval $i$. Then the order is approximated as:
\begin{equation}
    \label{eqorder}
    O=\frac{\underset{i}{\max} \hspace{.2cm}  h_{u_i}}{\sum_{i=1}^{\text{N}}h_{u_i}}+\frac{\underset{i}{\max} \hspace{.2cm}  h_{y_i}}{\sum_{i=1}^{\text{N}}h_{y_i}}
\end{equation}
The complexity is defined using the number of unique pitch values $U$ and the Shannon entropy of the interval distribution $E$ as: 
\begin{equation}
    \label{eqcomplexity}
    C=U(1+E)
\end{equation}
where the Shannon entropy is computed using the normalized pitch interval histogram $g_{i}=\frac{  h_{u_i}}{\sum_{i=1}^{\text{N}}h_{u_i}}$ as:
\begin{equation}
    \label{eqshannon}
    E=-\sum_{\{i:g_{i}>0\}} g_{i} \hspace{0.1cm} log_{2} \hspace{0.1cm} g_{i}
\end{equation}

The advantages offered by PCM include the ability to formulate the rules for music generation as constraints of the optimization problem in Eq. (\ref{eqnpcm}), as well as the effective incorporation of control aspects. However, PCM requires careful selection of the assessment function and its parameters to achieve realistic generation. More variables can be added to input and output for PCM, which leads to more realistic models, while the computation complexity of PCM increases. Next, we present two extensions of PCM, which are inspired by variants of MPC.

\subsection{Switched PCM}
Switched PCM uses switched system models for music generation \cite{Mhaskar2005}.  This is based on the switched MPC where the optimal switching index and control input are computed in a receding horizon manner by solving an optimization problem \cite{Zhang2016,Augustine2022}. Switched PCM can be implemented by identifying different models: $\textbf{f}_{\text{NN}_{\sigma}},$ $\sigma\in \{1,\dots, M\}$ where $M$ is the number of models or subsystems. Then during the PCM optimization problem the switching index sequence: $\bm{\sigma}_{k}=\left[\begin{matrix} \sigma_{k|k} & \sigma_{k+1|k} & \dots & \sigma_{k+\text{N}-1|k}\end{matrix}\right]^{\top}$ and control sequence $\textbf{U}_{k}$ are computed by solving the  optimization problem for switched PCM:
\begin{equation}
\label{eqnswitchedpcm}
    \begin{aligned}    \underset{\bm{\sigma}_{k},\mathbf{U}_{k},\mathbf{Y}_{k+1}}{\inf} \hspace{.2cm} &-A\\ \mathrm{subject \hspace{0.1cm}to}
    ~&~  \mathbf{Y}_{k+1}\in \mathbb{Y}^{\mathrm{N}},\hspace{0.2cm} \mathbf{U}_{k} \in \mathbb{U}^{\mathrm{N}}\\
    ~&~ \mathbf{y}_{i+1|k}=\mathbf{f}_{\text{NN}_{\sigma_{i|k}}}(\mathbf{y}_{i|k},\mathbf{u}_{i|k}) \hspace{0.7cm}k\in \mathbb{Z}^{+},i=k,...,k+\mathrm{N}-1.
    \end{aligned}    
\end{equation}
where the constraints are defined by the current active model or subsystem decided by $\sigma_{i|k}.$ The switched PCM can be used to introduce sudden variations in pitch or velocity, as well as to switch from one genre to another.
\subsection{Tube PCM}
Tube PCM uses the tube MPC \cite{Langson2004} principles for music generation. Tube MPC is a robust MPC approach where the output is constrained in a tube around the desired trajectory \cite{Langson2004}. In music generation, this idea can be used for generating variations or cover versions of a reference track. The resultant optimization problem for tube-PCM becomes:
\begin{equation}
\label{eqntubepcm}
    \begin{aligned}    \underset{\mathbf{U}_{k},\mathbf{Y}_{k+1}}{\inf} \hspace{.2cm} &-A\\ \mathrm{subject \hspace{0.1cm}to}
    ~&~  \mathbf{Y}_{k+1}\in \mathbf{Y}_{\text{r}_{k+1}} \oplus \mathbb{T}_{\mathbb{Y}} ,\hspace{0.2cm} \mathbf{U}_{k} \in \mathbf{U}_{\text{r}_k}\oplus\mathbb{T}_{\mathbb{U}}\\
    ~&~ \mathbf{y}_{i+1|k}=\mathbf{f}_{\text{NN}}(\mathbf{y}_{i|k},\mathbf{u}_{i|k}) \hspace{0.7cm}k\in \mathbb{Z}^{+},i=k,...,k+\mathrm{N}-1.
    \end{aligned}    
\end{equation}
where $\mathbb{T}_{\mathbb{Y}}\subset \mathbb{Y}$ and $\mathbb{T}_{\mathbb{U}}\subset \mathbb{U}$ are the tube sets for the output and input and $\oplus$ denotes the Minkowski sum. Note that compared to Eq. (\ref{eqnpcm}), only the constraint set for outputs and inputs is changed in tube PCM.

    


\section{PCM: Numerical Illustration}
This section presents a numerical example to illustrate the proposed PCM approach. As a first step, the score-rating NN is trained as described in Section \ref{secpcm}. The rating NN uses a single hidden layer containing 40 neurons with a sigmoid activation function. The training dataset consists of $\text{H}=100$ scores, as discussed in Section \ref{secpcm}. 
The mean squared error (MSE) for the training phase is obtained as $\text{MSE}_{\text{NN}}=\frac{\text{L}_{s}}{\text{H}}=0.0067$. In the next step, an RNN model is identified, with the model parameters selected as $L=2$, $n_{1}=30$. The training input comprises $\text{D}=100$ samples of MIDI note numbers, and the corresponding training output consists of the velocity values of the track. The framework can be extended to include additional input and output variables, resulting in a multivariable model. The RNN model is trained by minimizing the loss function given in Eq. (\ref{eqjyrnn}). The MSE of the training error is obtained as $\text{MSE}_{\text{RNN}}=\frac{\text{L}}{\text{D}}=0.0173$.

Using the RNN model as in Eq. (\ref{eqyhatrnn}) and rating NN-based loss function as in Eq. (\ref{eqjyrnn}), the PCM optimization problem is formulated as in Eq. (\ref{eqnpcm}), which is solved using the fmincon function in MATLAB. The parameters for PCM are chosen as $\text{N}=20,$ $\alpha_{s}=0.1$ and the control constraints are defined as $u_{min} \leq u_{k} \leq u_{max}$ with $u_{min}=60$ and $u_{max}=72$. The PCM optimization problem is solved for $\text{N}_{s}=20$ instants, and the first element of the input vector is stored in $\textbf{U}$. This resulted in a score with 20 notes for which the corresponding audio signal is generated as discussed in Section \ref{secmusicgen}, where the note duration is chosen as $d=0.5$. The parameters for the ADSR filter are chosen as $t_{A}=0.05d,$ $t_{D}=0.15d,$ $t_{S}=0.55d,$ $t_{R}=0.25d$ and $S=0.75$. 
\par
\begin{figure}[H]
 		\begin{center}
 		\includegraphics [scale=0.225] {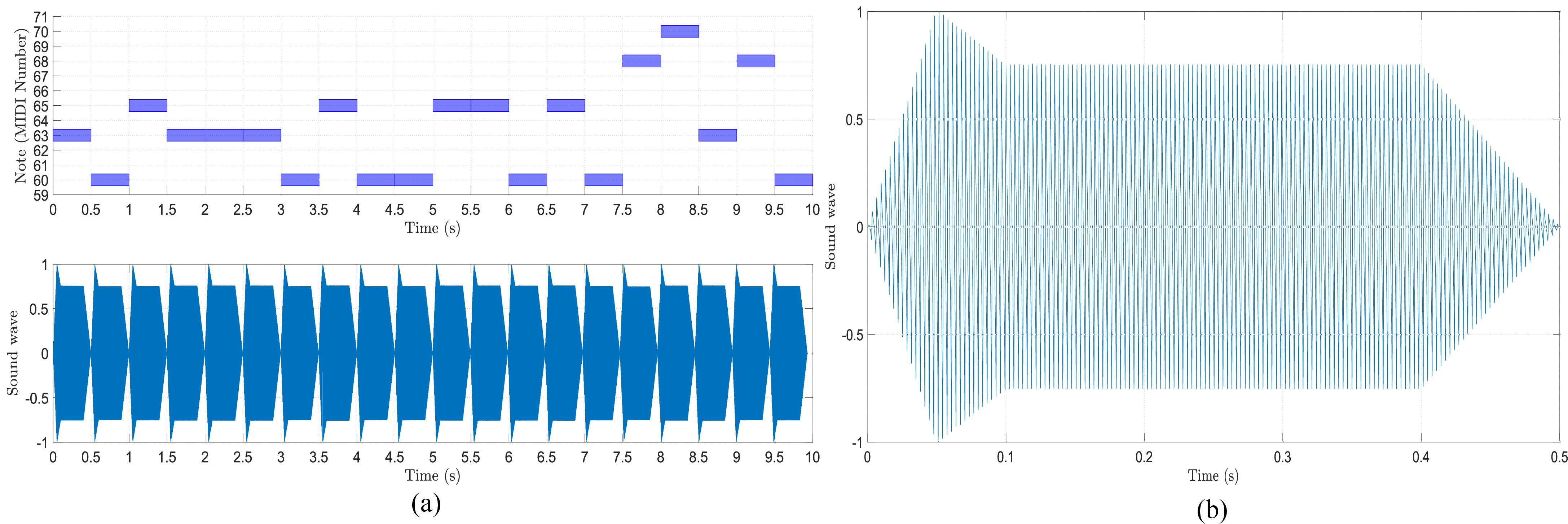}
 		\caption{(a) Score and sound wave  (piano) \hspace{0.3cm} (b) Zoomed sound wave.}
   \label{fig3}
 	\end{center}
 \end{figure}

 Fig. \ref{fig3}(a) shows the notes and sound wave generated by the proposed PCM. The sound wave for the first note is shown in Fig. \ref{fig3}(b), where the envelope generated by the ADSR filter is clearly visible. In this simulation, synthesis of a monophonic musical piece is considered with piano only. One can also add more accompaniments to the piano, such as drums, guitar, etc., to generate a polyphonic music which changes the shape of the sound wave according to the synthesis. 

\par
\begin{figure}[H]
 		\begin{center}
 		\includegraphics [scale=0.225] {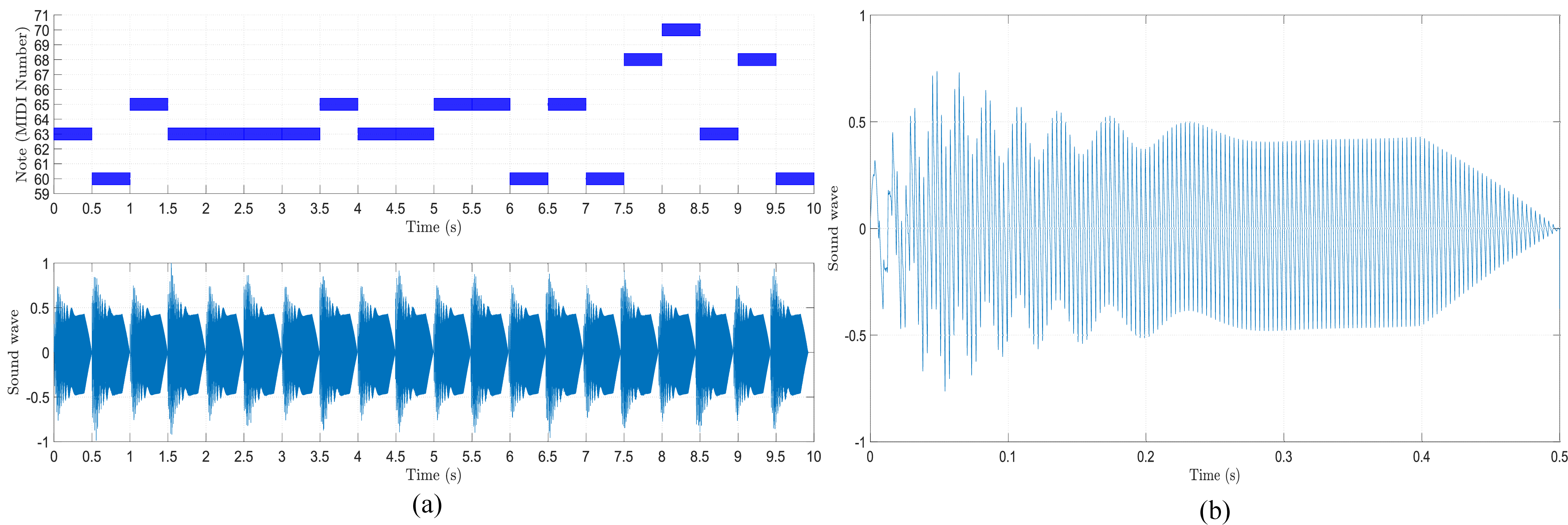}
 		\caption{(a) Score and sound wave  (piano and drums) \hspace{0.3cm} (b) Zoomed sound wave.}
   \label{fig4}
 	\end{center}
 \end{figure}

 Fig. \ref{fig4} shows the score and sound wave generated by the proposed PCM using a different RNN model, which was identified with a slightly different training dataset. During the synthesis stage, the piano is accompanied by drums. In this process, the sound waves for the piano and drum are generated separately, and the final sound is obtained by mixing both waves using the relation $x=x_{piano}+0.7x_{drum}$. The resulting sound wave is plotted in Fig. \ref{fig4}(b), where it can be observed that the envelope of the wave changes due to the addition of the drum. The sampling frequency for the sound wave is set to 44.1 kHz, and the resulting digital music for both Figs. \ref{fig3} and \ref{fig4} are stored as WAV files, which can be downloaded from the link \footnote{https://github.com/MIDHUNTA30/PCM}. 

\section{Conclusion}

This paper presented a controlled music generation approach called predictive controlled music (PCM), which is based on optimal control principles. The proposed PCM approach generates the notes by solving an MPC optimization problem in a receding horizon way. The PCM can incorporate the rules of music generation as constraints of the optimization problem in addition to a model extracted from the training data. This resulted in a control-based generation of music, which is illustrated using a numerical simulation. Formulations for different extensions of PCM, such as tube PCM and switched PCM, are also presented. The future works include analyzing stability and convergence properties of PCM.

\bibliographystyle{unsrt}

\end{document}